\documentclass[doublespacing,authoryear,12pt]{elsart}
\journal{Icarus}

\usepackage{natbib}
\usepackage{lscape}
\usepackage{graphicx}
\begin{document}
\begin{frontmatter}
\title{Conjugate observations of Saturn's northern and southern H$_3^+$ aurorae}

\author[UoL]{James O'Donoghue}, 
\author[UoL]{Tom S. Stallard},
\author[UoL]{Henrik Melin}, 
\author[UoL]{Stan W.H. Cowley}, 
\author[UoL]{Sarah V. Badman},
\author[BU]{Luke Moore},
\author[UCL]{Steve Miller},
\author[EP]{Chihiro Tao},
\author[JPL]{Kevin H. Baines}, and
\author[UoL]{James S.D. Blake}
\address[UoL]{Department of Physics and Astronomy,
University of Leicester, Leicester, LE1 7RH, (UK)}
\address[UCL]{Atmospheric Physics Laboratory, Department of Physics and Astronomy, University College London, London, WC1E 6BT, (UK)}
\address[JPL]{NASA Jet Propulsion Laboratory, M/S 183-601, 4800 Oak Grove Drive, Pasadena, CA 91109, (USA)}
\address[BU]{Center for Space Physics, Boston University, Boston, MA 02215 (USA)}
\address[EP]{Laboratoire de Physique des Plasmas, Ecole Polytechnique,
4 avenue de Neptune, 94100, Saint-Maur-des-Fosses, (France)}

\begin{abstract}
We present an analysis of recent high spatial and spectral resolution ground-based infrared observations of H$_3^+$ spectra obtained with the 10-metre Keck II telescope in April 2011. We observed H$_3^+$ emission from Saturn's northern and southern auroral regions, simultaneously, over the course of more than two hours, obtaining spectral images along the central meridian as Saturn rotates. Previous ground-based work has derived only an average temperature of an individual polar region, summing an entire night of observations. Here we analyse 20 H$_3^+$ spectra, 10 for each hemisphere, providing H$_3^+$ temperature, column density and total emission in both the northern and southern polar regions simultaneously, improving on past results in temporal cadence and simultaneity. We find that: 1) the average thermospheric temperatures are 527$\pm$18 K in northern Spring and 583$\pm$13 K in southern Autumn, respectively; 2) this asymmetry in temperature is likely to be the result of an inversely proportional relationship between the total thermospheric heating rate (Joule heating and ion drag) and magnetic field strength - i.e. the larger northern field strength leads to reduced total heating rate and a reduced temperature, irrespective of season, and 3) this implies that thermospheric heating and temperatures are relatively insensitive to seasonal effects.

\end{abstract}

\begin{keyword}
Saturn \sep aurora \sep magnetosphere \sep ionosphere \sep atmosphere
 

\end{keyword}
\end{frontmatter}
\section{Introduction}
Saturn's upper atmosphere is defined as the region at which the neutral molecular species within cease to mix convectively and are thus able to separate according to their scale heights; it is dominated by H, H$_2$, and He and the boundary between the mixing and non-mixing regions is known as the homopause \citep{2009nagy}. Within the upper atmosphere, solar radiation, particularly extreme ultraviolet (UV) is responsible for the production of the Saturnian ionosphere \citep{2009Moore}. In addition to this, the impact of electrons in the polar regions (accelerated along magnetic field lines), is responsible for the creation of the Saturnian aurora \citep[e.g.,][]{lamy2009uvskr}. Study of the Saturnian aurorae is generally divided into two wavelengths, the UV and infrared (IR). The former reveals the impact of highly energetic electrons on the polar ionosphere through the excitation of H and H$_2$ \citep{1983shemansky,gustin09}, whilst the latter is the focus of this study and is observed primarily via the discrete ro-vibrational emission lines of the molecular ion H$_3^+$. First detected on Saturn in late 1992 \citep{1993geb}, H$_3^+$ has served as a useful probe for examining the conditions in the ionosphere of Saturn. H$_3^+$ is produced indirectly from the ionisation of H$_2$, driven in the polar regions by particle precipitation at the boundary between open and closed planetary magnetic field lines \citep{Cow04open}, statistically located near $\sim$15$^{\circ}$ co-latitude \citep{badmansatwid}. Since H$_3^+$ is quasi-thermalized in the upper atmospheres of the giant planets \citep{1990mill}, examining emissions in the polar regions thus allows for the study of both the interactions between the ionosphere and its immediate space environment, and the physical conditions in the surrounding neutral atmosphere. \\
Prior to the findings of this study, measurements of auroral thermosphere temperatures have been limited to time-averaged values as Saturn's auroral H$_3^+$ emissions in the infrared are relatively weak: they are less than a hundredth the intensity of those at Jupiter when observed from Earth owing to Jupiter's higher ionospheric temperature \citep[$\sim$1000 K,][]{lam97} compared to Saturn's as we shall see, and this rising temperature causing an almost exponential increase in H$_3^+$ emission \citep[see][Figure 2]{96line}. \citet{melin07a} performed the most recent ground-based study of auroral thermospheric temperatures at Saturn, combining results from the 3.8-metre UKIRT telescope and CGS4 spectrograph obtained in 1999 and 2004, with exposure times of 210 and 26 mins, to derive a temperature of 450$\pm{50}$ K for the southern spring/summer auroral thermosphere. More recently, data from the Visual and Infrared Mapping Spectrometer (VIMS) on board the Cassini Saturn orbiter \citep{vims}, analysed by \citet{tssras} and \citet{melin11}, showed temperatures in the southern auroral region of 560-620$\pm{30}$ K over a 24-hour period in June 2007, and 440$\pm{50}$ K in measurements in September 2008, respectively. These temperatures are far warmer, by several hundreds of Kelvin, than predicted by models using only the Sun as an energy input \citep{YM}. We shall not heavily dwell on the longer term (months and years) effects here, except to say that both the IR and UV components of auroral emissions are modulated by solar wind conditions which do vary on such time scales, a fact owing to the variable solar wind dynamic pressure exerted on Saturn's magnetosphere \citep[see][and references therein]{kurthbook}.\\

Heating in the auroral region is thought to be dominated by Joule heating and ion drag via ionospheric Pedersen currents \citep{Cow04open,2005smith,aurtrans,2011galand,ingo2012}. The total heating from these two mechanisms generates $\sim$5 TW of power per hemisphere, whilst auroral particle precipitation provides an additional energy input at $\sim$0.1 TW \citep{Cow04open}. Only through a greater understanding of the mechanisms and conditions that cause physical parameters such as temperature, column density, and the total emitted energy over all wavelengths (henceforth, total emission) to persist or vary, can we start to add constraints to models and theories of the ionosphere. However, whilst individual temperature measurements have been made over long time scales, a study in both hemispheres simultaneously has not yet been performed. Here, we present and discuss the results of observations with the Near-Infrared SPECtrometer (NIRSPEC) \citep{nirs} instrument in high-resolution mode using the 10-metre W.M. Keck telescope situated on Mauna Kea, Hawaii. We study the main auroral region H$_3^+$ temperature, column density and total emission in the northern and southern hemispheres of Saturn at the same time with a temporal resolution of 15 minutes, and explore the implications of these measurements.
\section{Observations}
This study examines 132 minutes of observations obtained on 17 April 2011 using the 10-metre Keck telescope. These data were obtained using the NIRSPEC \citep{nirs} in high-resolution cross-dispersed mode with a resolution of R = $\lambda$/$\Delta\lambda$ $\sim$25,000, providing a minimum resolution of $\Delta\lambda\approx$ 1.59 x 10$^{-4}$ $\mu$m at 3.975 $\mu$m. The wavelength range used in this study is between 3.95 and 4.0 $\mu$m, covering the Q-Branch ($\Delta$J=0) ro-vibrational transition lines of H$_3^+$. The slit of the spectrometer was orientated in a north-south direction on Saturn aligned along the rotational axis, which we note is also co-aligned with the magnetic axis to within measurement uncertainties of $\sim$0.1$^{\circ}$ \citep{satintfield}. The planet is then seen to rotate beneath the slit allowing the acquisition of spectral images at a fixed local time of noon, but with a varying Saturn System III Central Meridian Longitude (CML). We were able to collect data between $\sim$10$^{\circ}$ and $\sim$22$^{\circ}$ co-latitude in both the northern and southern hemispheres given the viewing geometry at the time. The slit measures 0.432$^{\prime\prime}$ width by 24$^{\prime\prime}$ length with a pixel on the CCD corresponding to 0.144$^{\prime\prime}$ squared on the sky. The atmospheric seeing during this period was $\sim$0.4$^{\prime\prime}$. This dataset was recorded between 10:33:42 and 12:46:28 Universal Time (UT), covering a CML on Saturn of 103-176$^{\circ}$. Each set of spectra taken consist of twelve 5-s integrations, creating exposures 60 s long, consisting of object, or Saturn, `A' and sky `B' frames with the telescope slewing between the relevant positions of each in the sky in an ABBA pattern. An example of a typical 60 second `A-B' frame exposure (unreduced apart from the sky subtraction) is shown in Figure 1. The reduced spectral images are co-added over 15 minute segments to improve the signal to noise (S/N) ratio (this is elaborated on in the Data reduction section). During one such segment, Saturn rotates through 8.5$^{\circ}$ of longitude. On 17 April 2011, Saturn's northern hemisphere was tilted towards the Earth with a sub-Earth latitude of 8.2$^{\circ}$: in conditions of Saturn's northern spring. 
\begin{figure}
\centerline{\includegraphics[width=16.5cm]{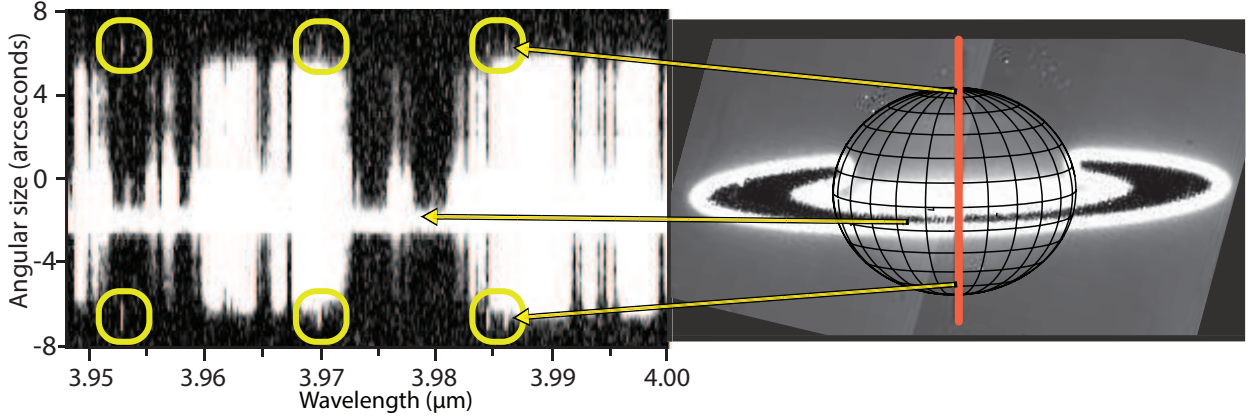}}
\caption{A typical sky-subtracted (A-B) spectral image of Saturn taken at Saturn local noon. The wavelength range is shown on the horizontal axis and the angular size in the sky is shown on the vertical axis. A real image of Saturn is shown to the right (taken with the Keck II telescope guide-camera) with arrows indicating the position of the aurorae and rings. Discrete H$_3^+$ emission lines can be seen as white vertical lines within the indicated yellow boxes. Hydrocarbon absorption of solar radiation appears as black between the auroral regions regions; hydrocarbons follow the general formula C$_n$H$_m$ where n and m are integer numbers, for example, methane is CH$_4$. The white bar of emission at -2$^{\prime\prime}$ is the continuum reflection of sunlight from the rings. The remaining white pixels are due to reflected sunlight.}
\end{figure}
\\
\section{Data reduction and analysis}
We examined the data summed between $\sim$10$^{\circ}$ and $\sim$22$^{\circ}$ co-latitude in the northern and southern hemispheres as stated, we found that the errors in parameters such as temperature were too large to perform a study of small-scale ($\sim$500-1000's km) structures known to exist within, for example, in the UV aurorae \citep{2011grod,mere13}. The error in assigning pixels to co-latitudes (pixel registration) was based on fitting a Gaussian curve to the large and very bright rings to find an exact latitude on the planet and then mapping poleward from that position to the planet's limbs, giving a negligible error of $\textless$0.1 pixel for each spectral image. The random error introduced by atmospheric seeing is $\sim$3 pixels, such that the pixel rows containing the main auroral emissions (which are well constrained in position), is able to receive light from adjacent pixels. This `smearing' of data becomes more prevalent towards the poles, leading to $\sim$3-7 degrees of smear for the auroral regions under study. In addition, the main auroral regions themselves vary in width and location within the aforementioned 12$^{\circ}$ co-latitude range. Standard sky subtraction and flux calibration (using the star HR 6035) techniques were applied to the data, accounting for the Earth's atmosphere, and correlating CCD count to physical photon flux. The methane (CH$_4$) hydrocarbon present in Saturn's upper atmosphere acts to both reflect and absorb solar radiation, depending on the wavelength of light. In Figure 1, sunlight is strongly reflected at some wavelengths, but CH$_4$ acts to absorb light at other wavelengths. Due to the increased column depth of CH$_4$ towards the limb, sunlight is absorbed enough to see auroral H$_3^+$ emission with little interference. However, as the reflected sunlight increases equatorward it eventually swamps the signal from most discrete H$_3^+$ emission lines, the low-latitude H$_3^+$ emissions are discussed in \citet{paper1}. The co-latitude range we selected led to a S/N ratio of $\sim$10-30, which is appropriate to give an uncertainties of between 5 and 10\%. Reflected sunlight is removed from the auroral emission by measuring the equatorial solar spectrum then subtracting an empirically calculated proportion of it from our auroral spectra. With this subtraction, the only significant spectral signature that remains is the auroral H$_3^+$ emission. \\
For a given temperature, H$_3^+$ produces a unique spectrum, such that there is a fixed temperature-dependent ratio between emission lines at different wavelengths. The H$_3^+$ temperature, T(H$_3^+$), is found using a fitting routine which uses the spectroscopic line list from \citet{96line} and the latest H$_3^+$ partition function constants from \citet{2010Miller} to produce expected transition-line intensities for a given temperature, this is also described by \citet{Melin2013u} in further detail. The spectral function of H$_3^+$ is varied until the line-ratios match the least squares fit to the observed data, as shown by the 15 minute-integrated spectrum in Figure 2. For a given temperature the emission from a single H$_3^+$ molecule is known, so by dividing the observed emission by the molecular emission, we can find the number of emitting molecules, i.e. the H$_3^+$ column density. We produce a line-of-sight corrected column density, N(H$_3^+$), by measuring the differing observed path-lengths through the atmosphere across the disk of the planet. This varies as the sine of the colatitude as observed from Earth, accounting for the varying line-of-sight column depth as it increases towards the poles, e.g. multiplying N(H$_3^+$) by the sine of the colatitude in which it is located, will reduce the observed density to create a column density. Saturn's sub-Earth latitude was 8.2$^{\circ}$, by the addition of this angle to the observed latitude we correctly locate the position of pixels, e.g. at 80$^{\circ}$ north from Earth's perspective is actually 88.2$^{\circ}$ on Saturn. A measure of wavelength-integrated total emission from a line-of-sight corrected column of H$_3^+$, E(H$_3^+$) can then be calculated by multiplying the total emission per molecule by N(H$_3^+$). This parameter was introduced by \citet{lam97} for Jupiter, to be used as a separate parameter for studying the ionosphere, due to their observation of an anti-correlation between temperature and column density.  It is important to note that whilst the temperature is measured using the relative amplitudes of the various spectral peaks, the density is derived subsequent to the determination of temperature. Therefore, even though the least-squares fitting routine described above was used, the derived T(H$_3^+$) and N(H$_3^+$) are relatively independent parameters. E(H$_3^+$) is a useful parameter as it reveals the amount of energy lost by the ionosphere via radiative cooling to space. Throughout this paper the errors shown for the above parameters are $\pm$1 standard deviation (1-sigma), these arise from the uncertainty in the Gaussian fits to each spectral line (see Figure 2).   \\
\begin{figure}
\centerline{\includegraphics[width=13cm]{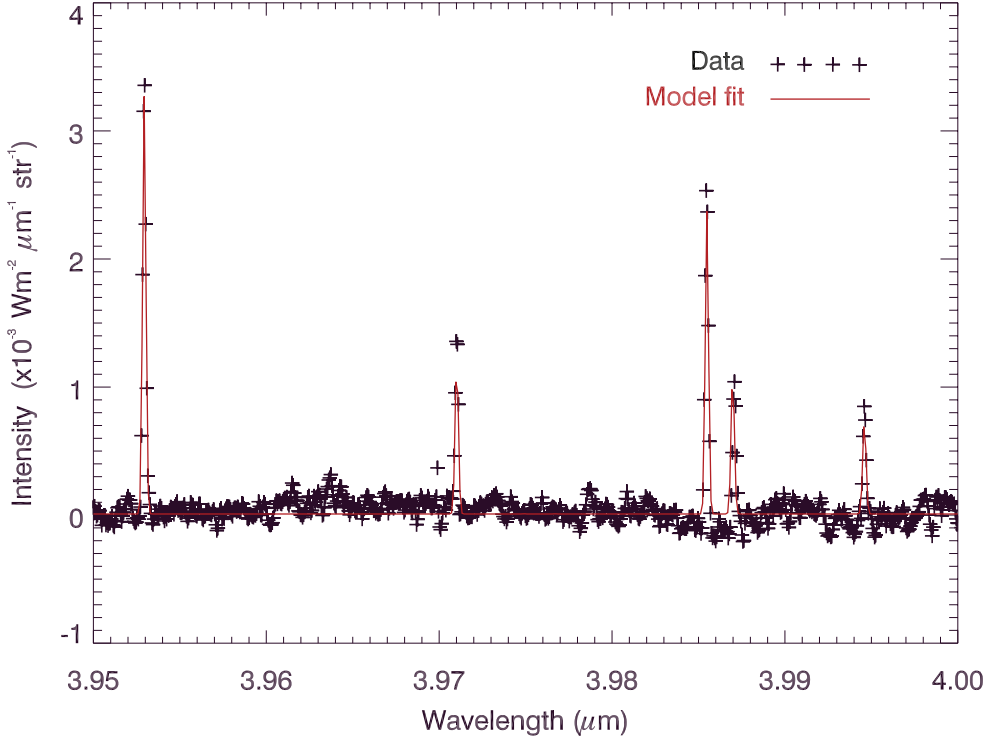}}
\caption{An example model fit to the data. H$_3^+$ emission is shown as a function of wavelength (black crosses), fitted to a model of the expected emission (red) for a given temperature. The data in this figure are from a 0.144$^{\prime\prime}$ (height) by 0.432$^{\prime\prime}$ (width) area of the planet, which corresponds to an area of approximately 3320 km x 2700 km at 19$^{\circ}$ co-latitude. From this fit we obtained a temperature of 523$\pm$13 K. The S/N ratio is $\sim$25 in this typical spectral profile. Low levels of solar reflection from hydrocarbons are visible over all wavelengths, though these levels are much reduced from their original values by the empirically calculated solar reflection subtraction.}
\end{figure}
\section{Results and discussion}
In Figure 3 we present simultaneous measurements of Saturn's H$_3^+$ temperature, column density, and total emission in the northern and southern auroral regions at local noon as a function of time. We view the aurorae as they rotate past the spectrometer slit and so variability is a combination of temporal changes occurring at local noon and longitudinal variations rotating into view. The relationships between the temperature, column density, and total emission between hemispheres are investigated. A summary of the results is also given in Table 1. The northern thermospheric temperature is on average 527$\pm$18 K, while the southern is 583$\pm$13 K. The column density averages for the north and south aurorae are 1.56$\pm$0.32 x10$^{15}$ m$^{-2}$ and 1.16$\pm$0.14 x10$^{15}$ m$^{-2}$, respectively. An anti-correlation between H$_3^+$ temperature and column density is observed in our data. The total emission is $\sim$1.5 times higher in the south, 0.98$\pm$0.02 Wm$^{-2}$str$^{-1}$, compared with 0.65$\pm$0.03 Wm$^{-2}$str$^{-1}$ in the north. This result is similar to previous work based on Cassini VIMS observations in which they examined IR wavelengths associated with H$_3^+$ emission at $\sim$3.6  $\mu$m, which showed the pre-equinox southern main region to be on average $\sim$1.3 times more intense than the northern main auroral region \citep{2011badman}. The higher levels of emission cause the S/N ratio in the south is$\sim$30 whilst the north is $\sim$18 on average, so that the errors in all parameters are lower there relative to the north (see again Table 1). 
\begin{landscape}
\begin{table}
\caption{Saturn's main auroral region properties as a function of time. Data obtained on 17 April 2011. All uncertainties shown are one standard deviation (i.e. 1-sigma errors).}
\begin{tabular}{c c c c c c c c}
Start Time & E(North,H$_3^+$)  &  T(North,H$_3^+$) &  N(North,H$_3^+$)  &  E(South,H$_3^+$) & T(South,H$_3^+$) & N(South,H$_3^+$)\\
(UT) & (x10$^{-5}$ Wm$^{-2}$sr$^{-1}$) &  (Kelvin) & (x10$^{15}$ m$^{-2}$) & (x10$^{-5}$ Wm$^{-2}$sr$^{-1}$) &  (Kelvin) & (x10$^{15}$ m$^{-2}$) \\ 
\hline
$10:33$  & 0.81 $\pm$0.03 & 528 $\pm$17 & 1.85 $\pm$0.34 & 0.92 $\pm$0.02 & 592 $\pm$13 & 0.96 $\pm$0.11 \\
$10:49$  & 0.69 $\pm$0.02 & 529 $\pm$17 & 1.59 $\pm$0.30 & 0.90 $\pm$0.02 & 580 $\pm$14 & 1.10 $\pm$0.14 \\
$11:02$  & 0.64 $\pm$0.02 & 539 $\pm$18 & 1.33 $\pm$0.25 & 0.89 $\pm$0.02 & 575 $\pm$14 & 1.16 $\pm$0.15 \\
$11:19$  & 0.66 $\pm$0.03 & 544 $\pm$19 & 1.29 $\pm$0.25 & 0.96 $\pm$0.02 & 588 $\pm$14 & 1.07 $\pm$0.13 \\
$11:36$  & 0.64 $\pm$0.03 & 537 $\pm$19 & 1.41 $\pm$0.28 & 1.00 $\pm$0.02 & 586 $\pm$13 & 1.15 $\pm$0.13 \\
$11:49$  & 0.61 $\pm$0.03 & 528 $\pm$20 & 1.49 $\pm$0.33 & 1.06 $\pm$0.02 & 585 $\pm$13 & 1.22 $\pm$0.14 \\
$12:04$  & 0.57 $\pm$0.03 & 513 $\pm$20 & 1.66 $\pm$0.39 & 1.06 $\pm$0.02 & 580 $\pm$12 & 1.30 $\pm$0.20 \\
$12:17$  & 0.60 $\pm$0.03 & 521 $\pm$20 & 1.51 $\pm$0.35 & 1.02 $\pm$0.02 & 579 $\pm$13 & 1.25 $\pm$0.14 \\
$12:32$  & 0.62 $\pm$0.02 & 517 $\pm$18 & 1.66 $\pm$0.34 & 0.99 $\pm$0.02 & 580 $\pm$13 & 1.19 $\pm$0.14 \\
$12:46$  & 0.65 $\pm$0.02 & 515 $\pm$15 & 1.77 $\pm$0.32 & 0.97 $\pm$0.02 & 579 $\pm$12 & 1.20 $\pm$0.13 \\
\end{tabular}
\end{table}
\end{landscape}
\begin{figure}
\noindent\includegraphics[width=34pc]{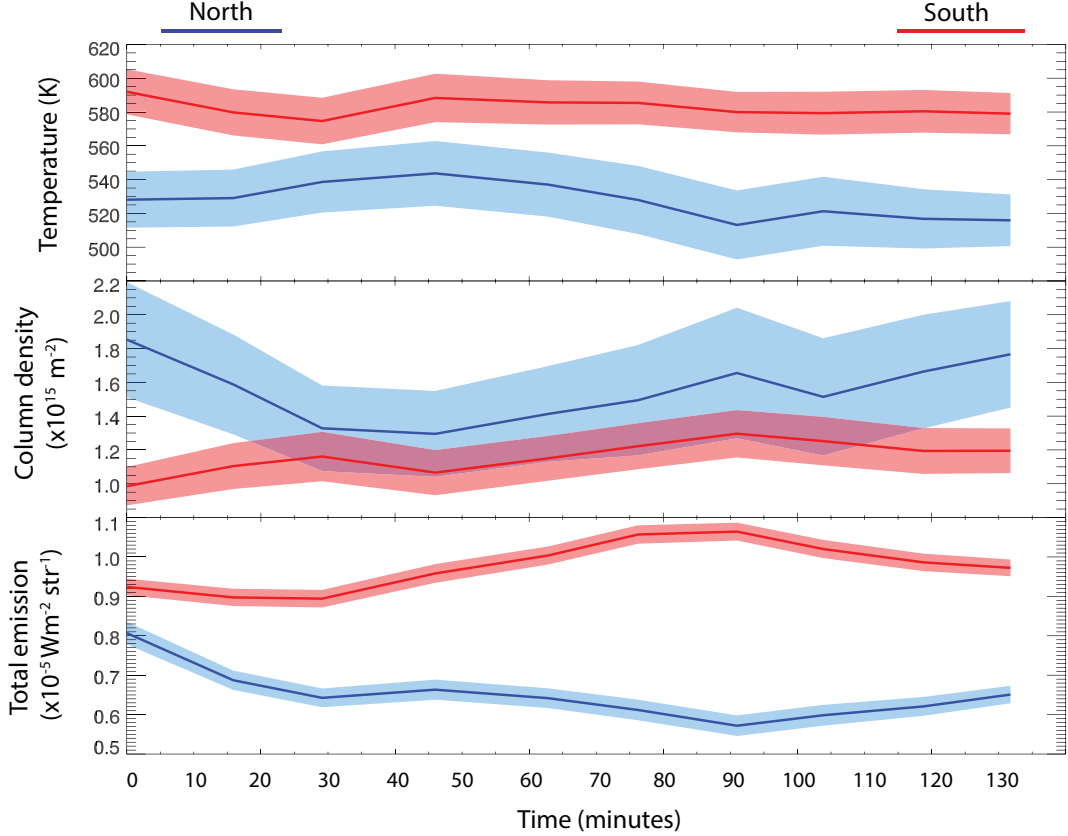}
\caption{Temperature, H$_3^+$ column density and total emission of Saturn's main auroral region emissions for each hemisphere integrated over 10-22$^{\circ}$ co-latitude (y-axis), plotted as a function of time (x-axis). Northern data are shown in blue, and southern data in red. The thin dark coloured lines show data values, while the light coloured shading shows their corresponding uncertainty ranges. The error bars show uncertainties of one standard deviation (i.e. 1-sigma error bars). The time at 0 minutes is 10:34 UT on 17 April 2011.}
\end{figure}

\subsection{Interhemispheric asymmetry in temperature and emission}
The average auroral thermospheric temperature in the south, $\sim$583 K, is within the 560-620 K range found by \citet{tssras} using Cassini VIMS data from June 2007. However, this is substantially higher than the ground-based 1999/2004 UKIRT result of 400$\pm$50 K found by \citet{melin07a} and the Cassini VIMS September 2008 result of 440$\pm$50 K by \citet{melin11}. These differences suggest that while temperatures are stable on the short time scales observed here, highly variable auroral temperatures can be seen on longer time scales. The relatively higher temperatures in both hemispheres here may indicate that Saturn is in the midst of a slow `heating event' on time scales greater than hours. Such an event has been observed on Jupiter by \citet{aurtrans} and takes place over a period of 3 Earth days, but this heating event was due to the compression of closed field lines, such that we can expect the effects to be symmetric north and south. Only through further observations taken of the same latitudes can we identify the nature of such long-term trends. The most striking result shown in Figure 3 is that the southern auroral thermosphere is significantly hotter and more emissive than the north over the $\sim$2 hour duration of these observations. Although the observations represent a `snapshot' of the possible conditions in Saturn's ionosphere, the following discussions and conclusions assume this represents conditions that are typical on Saturn at that time in its season. Additional observations are required over time scales of weeks and months, to validate that this asymmetry is not due to short term (hours or days) effects. To investigate the reasons for this unexpected temperature difference, we consider the combined Joule and ion drag heating rate per unit area of the ionosphere, in particular the effect of the hemispheric difference in ionospheric magnetic field strength, where the northern polar field is a factor $\sim$1.2 times the strength of the southern polar field (both integrated between $\sim$10$^{\circ}$-22$^{\circ}$) due to the quadrupole term in the planet's internal field. This is illustrated in Figure 4, where we plot the field strength in the Pedersen layer versus co-latitude from the respective poles for the northern (solid line) and southern (dashed) hemispheres, respectively. Here we have used the latest internal field model based on Cassini data by \citet{satintfield}, consisting of axial dipole, quadrupole and octupole terms, evaluated at an altitude of 1000 km above the IAU 1 bar reference spheroid. The latter 1 bar surface has equatorial and polar radii of 60,268 and 54,364 km, respectively, with the Pedersen layer located $\sim$1000 km above \citep[e.g.,][]{satintfield}.
\begin{figure}
\noindent\includegraphics[width=33pc]{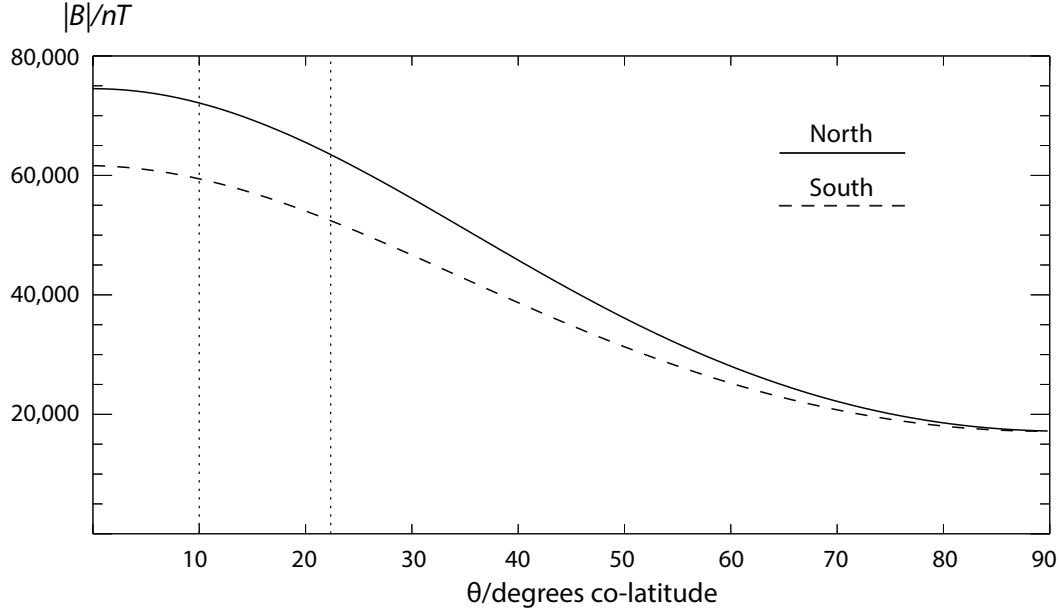}
\caption{Saturn's internal magnetic field strength $\mid$B$\mid$ in the ionospheric Pedersen layer using the \citet{satintfield} model, shown plotted as a function of planetocentric co-latitude in degrees from the corresponding pole for the northern (solid) and southern (dashed) hemispheres, respectively. The Pedersen layer is taken to lie 1000 km above the IAU 1 bar pressure reference spheroid. The vertical lines (dotted) indicate the range of auroral region co-latitudes.}
\end{figure}

The combined Joule and ion drag thermospheric heating rate per unit area in the northern (N) and southern (S) hemispheres is given by (e.g., \citet{2005smith,2005cow})
\begin{equation}
q_{N,S}=\Sigma^*_{P_{N,S}}E^2_{eqN,S}~, 
\end{equation}						

where $\Sigma^*_{P_{N,S}}$ is the effective height-integrated Pedersen conductivity of the ionosphere, modified from the true value $\Sigma_{P_{N,S}}$ due to drag-induced atmospheric sub-corotation, and $E_{eqN,S}$ is the equatorward-directed ionospheric electric field (\textbf{\textit{E}}= -\textbf{\textit{V}}$\times$\textbf{\textit{B}}) in the rest frame of the planet at a given co-latitude with respect to the rotation/magnetic axis. The latter is given by
\begin{equation}
  E_{eqN,S} = \rho_{iN,S}(\Omega_{Sat} - \omega ) B_{iN,S}~,  		
  \end{equation}	

where $\rho_{iN,S}$ is the perpendicular distance of the Pedersen conducting layer from the axis, $\Omega_{Sat}$ is the angular velocity of Saturn defining the planetary `rest frame' of rigid corotation, $\omega$ is the magnetospheric plasma angular velocity on the field line passing through the ionosphere at that co-latitude, and $B_{iN,S}$ is the corresponding ionospheric magnetic field strength. The field is taken to be uniform and perpendicular to the polar ionosphere to a sufficient approximation, the latter peaking in Pedersen conductivity at an altitude of $\sim$1000 km in the auroral region \citep{2011galand}. The effective Pedersen conductivity is given by
\begin{equation}
\Sigma^*_{P_{N,S}} = (1-k)\Sigma_{P_{N,S}}~, 						
\end{equation}	
where \textit{k} is the ratio between the neutral atmosphere angular velocity and the plasma angular velocity in the planet's frame
\begin{equation}
(\Omega_{Sat} - \Omega^*_{Sat} ) = k(\Omega_{Sat}-\omega)~,	
\end{equation}
where $\Omega^*_{Sat}$ is the angular velocity of the neutral atmosphere. Atmospheric modelling results indicate that \textit{k}$\sim$0.5 at Saturn \citep{2011galand}. Combining Equations (1) and (2) we obtain
\begin{equation}
q_{N,S} = \Sigma^*_{P N,S}~\rho^2_{iN,S}~(\Omega_{Sat}-\omega)^2~B^2_{iN,S}~. 			
\end{equation}
Now to the same order of field approximation (in which terms quadratic in the co-latitude are neglected relative to zeroth order) the magnetic flux threading the ionosphere between the pole and cylindrical distance $\rho_{iN,S}$ from the axis is
\begin{equation}
\Phi_i = \pi\rho^2_{iN,S}~B_{iN,S}~,				
\end{equation}
such that we can write Equation (5) as 
\begin{equation}
q_{N,S}= \frac{1}{\pi}~\Sigma^*_{P N,S}~(\Omega_{Sat}-\omega)^2~\Phi_iB_{iN,S}~.				
\end{equation}
If we then consider conjugate points in the northern and southern hemispheres joined by a common field line, thus containing equal magnetic flux $\Phi_i$ and having equal plasma angular velocities $\omega$, as required under steady state conditions, it can be seen that the relative heating rates per unit area north and south depend only on the product of the effective height-integrated Pedersen conductivity, $\Sigma^*_{P N,S}$, and the field strength in the ionosphere, $B_{iN,S}$. However, for approximately equal ionospheric Pedersen layer number densities north and south (as addressed in the next section), the Pedersen conductivity is expected to vary approximately inversely as the ionospheric field strength, as reported by \citet{2011galand} for near-equinoctial conditions. In these circumstances the thermospheric heating rates per unit area will be equal in the two hemispheres at conjugate points, independent of the magnetic field strength. This result does not therefore give immediate reason to expect the southern thermosphere to be hotter than the northern, unless the northern ionospheric conductivity is lower than that in the south by an unexpectedly large factor. 
We note, however, that the above result also implies that the total heat input to the thermosphere from Joule heating and ion drag integrated over the whole polar region will be larger in the south than in the north, because the area of heating is larger in the south than in the north due to the lower field strength. If we consider conjugate circular ionospheric strips north and south with equal magnetic flux $d\Phi_{i}=2\pi B_{iN,S}~\rho_{iN,S}~d\rho_{iN,S}$, then the total heating north and south in the strips is given by
\begin{equation}
dQ_{N,S}= 2\pi q_{N,S}~\rho_{i,N,S}~d\rho_{iN,S}= \frac{1}{\pi}\Sigma^*_{P N,S}~(\Omega_{Sat}-\omega)^2~\Phi_id\Phi_i~.
\end{equation}
Thus the total heating rate, obtained by integrating over all the flux strips from the pole to the point where rigid corotation is attained, is then proportional only to $\Sigma^*_{P N,S}$, such that if the latter varies approximately inversely with the field strength as indicated above, the total power dissipated to heat in the polar thermosphere will larger in the southern hemisphere than in the northern. It remains to be investigated by modelling whether such an effect could produce the temperature differences measured here.  If not, then some other heating mechanism, such as hemispheric differences in wave driving from below, must be implicated.

A comparison with recent modelling work by \citet{2011galand} agrees with the interpretation above in that an asymmetry is present (during equinox) in which the Pedersen and Hall conductivities were 1.2 and 1.3 times higher, respectively, in the southern hemisphere than in the northern. Previous observations by Cassini VIMS analysed by \citet{2011badman} also found the same trend in intensity - and likely therefore in temperature - in their pre-equinox 2006-2009 data. The fact that this asymmetry persists post-equinox in our data suggests that the magnitude of the magnetic field is the dominant effect on Pedersen conductivity rather than the solar extreme-UV ionization, at least in northern spring. In other words, magnetic field strength may dominate over seasonal effects in determining inter-hemispheric auroral thermosphere temperatures, though further observations and modelling are required to test this - in particular whether or not it persists into the northern summer season. \\
In the UV, simultaneous observations of the conjugate northern and southern aurora taken by the HST in 2009 have been analysed by \citet{UVaurora} and \citet{mere13}. The former showed, from the data acquired over a period of $\sim$1 month just pre-equinox, that the northern main auroral region had on average $\sim$17\% higher emitted power than the south, the opposite to the IR case presented here. The latter study found transient eastward-propagating patches of UV emission in the dawn-to-noon sector for 70\% of the 32 visits using the HST, these patches are similar to the small-scale features found by \cite{2011grodgrapes}. In this study, such small-scale features are therefore very likely to be passing by the spectrograph slit, and could lead to small-scale variations in the column density of H$_3^+$. However, UV emissions are a prompt emission in which hydrogen is excited by particle precipitation and immediately releases the newly acquired energy to space via the emission of UV photons. Hydrogen that emits in the UV is not therefore in thermal equilibrium with the surrounding thermosphere. By contrast, H$_3^+$ emission is largely driven by temperature changes due to Joule heating and ion drag, so to a good approximation H$_3^+$ is thermalized with the thermosphere \citep{millsept06}. As a consequence of the differing IR and UV emission production mechanisms, direct comparison is difficult. Given that Joule heating and ion drag is $\sim$50 times greater in power than auroral particle precipitation \citep{Cow04open} (responsible for UV emission), it is understandable in the above UV studies that a stronger northern UV emission or the appearance of small-scale structures/patches need not necessarily correspond to higher temperatures or IR emission. We were unable to resolve small-scale features here with small uncertainties, so we cannot compare individual features. In addition, it should be noted that the above UV observations took place over 2 years earlier than those presented here.

\subsection{Altitudinal considerations}
The ratio between the average column integrated densities in the north and south auroral regions is 1.35. Broadly speaking, the cause for this asymmetry could be an increase in the northern H$_2$ ionization rate, which itself arises from the larger incident solar photon flux in the north owing to Saturn's 9.1$^{\circ}$ sub-solar latitude. An increase in ionized H$_2$ then leads to a greater production of H$_3^+$. This was also demonstrated using a 1-D model, the Saturn Thermosphere Ionosphere Model (STIM), which was utilised by the work presented here and found a range of north-south H$_3^+$ density ratios between 1.2-3.8 based on solar EUV influx alone (between 10-22$^{\circ}$ co-latitude), for several different values of vibrationally excited H$_2$. \\
Despite having lower column densities, the average total emission is $\sim$1.5 times higher in the south, this is due to the higher temperature there. The total emission is on average not related to the column density as there are fewer H$_3^+$ molecules emitting and yet there is more emission, so there is clearly more emission per molecule because the emitting molecules are hotter. \\
One might assume that a higher H$_3^+$ density and lower temperature (as in the northern auroral thermosphere) could indicate that the column of H$_3^+$ sampled was deeper in the atmosphere: an inverse relationship like this exists at Jupiter \citep{lystrup8}. As previously stated, the H$_3^+$ densities presented herein are column integrated and line of sight corrected, such that altitudinal information is averaged for the observed atmospheric column. If the H$_3^+$ densities were higher in the northern hemisphere relative to the south because of the inverse relationship above, it implies that electrons must penetrate deeper in the north, thus leading to enhancements in H$_3^+$ production (density) in a colder region. For this to occur, the electron precipitation energy must be relatively higher in the northern auroral region, since higher energy electrons penetrate to lower altitudes than lower energy electrons \citep{2011Tao}. To test this, we employed the magnetosphere-ionosphere coupling model known as the `CBO' model, derived by \citet{2003cow} and \citet{Cow04open}, and used updated parameters derived from Cassini spacecraft measurements \citep{2008cow}. It is appropriate to reduce the northern conductivity from 4 mho in \citet{2008cow} by a factor of 1.215 corresponding to the field asymmetry already mentioned, such that it is fixed at 3.3 mho whilst the south remains at 4 mho. Following this, we find that the field-aligned current density, precipitating electron energy flux, and average electron precipitation energy in both hemispheres are closely similar, the latter parameter being $\sim$11.4 keV. Therefore, we have no reason to expect strong differences in the altitude at which auroral electrons are deposited in either hemisphere, nor the overall number density profile at those altitudes. This expectation is based on electrons accelerated planet-ward along closed field lines that require accelerating voltages of $\sim$10 kV to reach the ionosphere. Poleward of this, currents along open field lines can be carried by cool dense magnetosheath plasma that requires accelerating voltages of $\sim$0.1-1 kV \citep{08Bunceopen}, hence the majority of auroral emission is associated with particle precipitation along closed field lines.  \\
The previous section concerned itself with an inverse relationship between Pedersen conductivity and field strength, with the caveat being that the number density of the Pedersen layer is the same for both hemispheres, in line with the above expectation. In Figure 3 it appears at first glance that the contrary is true because H$_3^+$ density is higher in the north, which would imply our previous argument is incomplete. However, H$_3^+$ density which peaks at an altitude of 1155 km \citep{tssras} is not wholly representative of the Pedersen layer density which itself peaks at an altitude of 1000 km \citep{2010moore,2011galand}. The ions (and their companion electrons) that create the Pedersen layer are largely hydrocarbon ions, which are dominant below $\sim$1000 km \citep{ingo2012}, and so the H$_3^+$ density is neither a proxy for the conducting layer density and does not therefore have direct implications for the previous derivation. \\
Following from the above, the H$_3^+$ temperature must differ from the Pedersen layer temperature because it is higher in altitude in a region of positive temperature gradients. However, an increase in temperature at 1000 km will lead to an increase in temperature at altitudes above it due to the vertical conduction of heat. So, in contrast to the H$_3^+$ column density, the H$_3^+$ temperature is a useful proxy for the thermal conditions within the conducting layer. \\

\subsection{Correlations between parameters}
The northern auroral thermosphere exhibits an anti-correlation between temperature and column density of -0.72. This is less pronounced in the south, with a correlation coefficient of -0.52. These anti-correlations are based on small variations in these parameters - small because they remain within the errors bars of neighbouring values, such that variability is within the uncertainty, particularly for temperatures. Recent work by \citet{melinanti} shows that this is a physical phenomenon provided that the anti-correlation is outside of the range of uncertainties, as opposed to a product of the least-squares fit, i.e. as temperature increases, column density decreases and \textit{vice versa}. If the anti-correlation here is real, the physical ramification may be that increases in the density of H$_3^+$ lead to decreases in temperature, i.e. H$_3^+$ may be acting as a `thermostat' to cool the planet in a small way as it does on Jupiter and Uranus \citep{millsept06}, although recent work by \citet{ingo2012} shows such cooling plays a minor role at Saturn. 
In the north, there is a positive correlation coefficient of 0.54 between temperature and total emission whilst the correlation coefficient between total emission and column density is significantly weaker at 0.13, suggesting that it is changes in temperature that modulate changes in total emission. In the south, the H$_3^+$ column density and total emission are instead correlated strongly, with a coefficient of 0.76, contrasting with the north, though such correlations are the result of variability of similar size to the uncertainties involved. For both hemispheres, new parts of the main aurorae are passing by the spectrograph slit and there may also be there are large changes in the particle precipitation at local noon during this period. Both such effects have been observed in the H$_3^+$ aurora using Cassini VIMS data by \citet{badman2012a,badman2012b}. Due to the observational techniques employed here (observing only Saturn local noon), it is difficult to distinguish between these processes.

\section{Conclusions}
Ground based Keck NIRSPEC observations of auroral H$_3^+$ emission from Saturn have been analyzed. During the $\sim$2 hours of data obtained here, temperatures remained effectively constant within the error range achieved. At the same time, column density and total emission vary greatly over time in each hemisphere evidenced by Figure 3. This may be caused by temporal or spatial variation in the aurora, likely due to varying particle precipitation, leading to the small variations seen in all H$_3^+$ parameters. The main auroral region in the south is significantly warmer and more emissive than its northern counterpart. This asymmetry is attributed to an inversely proportional relationship between ionospheric Joule and ion drag heating and magnetic field strength. Somewhat unexpectedly, this effect outweighs the increased heating produced by seasonal enhancements in conductivity, meaning that the southern autumn aurora is 50 K warmer than that in the northern spring hemisphere. This is consistent with model predictions of a higher Pedersen conductivity in the south than the north \citep{2011galand} and an intensity asymmetry observed by Cassini VIMS pre-equinox \citep{2011badman}. A number of correlations exist between parameters that may be significant and we highlight possible causes for them. A dedicated observing campaign of Saturn's aurora is required to verify these relationships and assess the long-term behaviour of Saturn in response to seasonal variations. Although the southern aurora is unfortunately no longer viewable until at least 2023 from Earth-based telescopes, the changing viewing geometry will allow for more comprehensive studies of the northern aurora for several years.

\section{Acknowledgements}
The data presented herein were obtained at the W.M. Keck Observatory, which is operated as a scientific partnership among the California Institute of Technology, the University of California, and NASA. The observations were made to support the Cassini auroral campaign in April 2011. Discussions within the international team lead by Tom Stallard on `Comparative Jovian Aeronomy' have greatly benefited this work, this was hosted by the International Space Science Institute (ISSI). We would like to thank Marina Galand for modelling assistance and related discussions. The UK Science and Technology Facilities Council (STFC) supported this work through the PhD Studentship of J.O'D. and consolidated grant support for T.S.S., S.W.H.C. and H.M. whilst S.V.B. was supported by a Royal Astronomical Society Research Fellowship.
\bibliography{TIME2012bib}
\bibliographystyle{agufull08}
\end{document}